\documentclass[preprint]{aastex}

\slugcomment{Mar. 20. 2010 version}

\shorttitle{Sunyaev-Zel'dovich Effect in a Radio Galaxy}
\shortauthors{Yamada et al.}

\begin{document}

\title{Search for the Sunyaev-Zel'dovich Effect in a Giant Radio Galaxy B1358+305}

\author{Masako Yamada\altaffilmark{1}} 
\email{masako@asiaa.sinica.edu.tw}

\author{Yutaka Fujita\altaffilmark{2}}

\author{Hiroshi Matsuo\altaffilmark{3}}

\and

\author{Naoshi Sugiyama\altaffilmark{4,5}}
\affil{}

\altaffiltext{1}{Institute of Astronomy and Astrophysics, Academia Sinica, P. O. Box 23-141, Taipei, 
 10617, Taiwan, R.O.C.}
\altaffiltext{2}{Department of Earth and Space Science, Osaka University, Toyonaka, Osaka, 560-0043}
\altaffiltext{3}{Advanced Technology Center, National Astronomical Observatory of Japan, Mitaka, Tokyo, 181-8588}
\altaffiltext{4}{Department of Physics, Nagoya University, Furo-cho, Chikusa-ku
Nagoya, 464-8602}
\altaffiltext{5}{Institute for the Physics and Mathematics of the Universe, University of Tokyo,
Kashiwanoha, Kashiwa, Chiba 277-8568, Japan}

\begin{abstract}
We present results of an imaging observation of the central region of a giant radio galaxy B1358+305.
The classical, standard scenario of Fanaroff-Riley II radio galaxies suggests that shock produced
 hot electrons contained in a radio galaxy are a good reservoir of the
 jet-supplied energy from active nuclei.
The aim of our observation is to search for the Sunyaev-Zel'dovich effect induced by these hot electrons.
The observation was performed at 21 GHz with the Nobeyama 45-m telescope.
Deep imaging observation of a wide region of size
$6.7^{\prime}\times6.7^{\prime}$  with the beam size $\theta_{\rm
  HPBW}=81.2^{\prime\prime}$ enables the most detailed examination of
the possible thermal energy of electrons contained in a radio galaxy. 
The resultant intensity fluctuation is 0.56 mJy beam$^{-1}$ (in terms
of the  Compton $y$-parameter, $y=1.04\times 10^{-4}$) at a 95 \%
confidence level.  
The intensity fluctuation obtained with imaging analysis sets the most 
stringent upper limit on the fluctuations in the central region of a giant radio galaxy 
obtained so far, and our results will be a toehold for future plans of SZE observation in a radio galaxy.
\end{abstract}

\keywords{cosmic microwave background -- Cosmology: observations --
  Galaxies: active -- Galaxies: individual:B1358+305 -- Radio
  continuum: galaxies}

\section{Introduction}

Extended lobes of radio galaxies are interesting as a probe of the energetics of active galaxies.
In the standard model of radio galaxies of Fanaroff-Riley type II \citep[FR-II;][]{fanaroff1974}, radio
lobes are formed by shock interactions of the jets with the surrounding intergalactic medium at the jet extremities \citep{scheuer1974,blandford1974}.
In this scenario, radio lobes consist mainly of the jet-supplied matter that passed 
the termination shock. 
The gas density in the radio lobes is sufficiently low that radiative cooling is ineffective, 
resulting in the energy supplied to the radio lobes to be well conserved for the lifetime of the 
radio galaxy.
The lobe has a higher pressure than the ambient intergalactic medium (IGM) 
due to shock-compression, and it can expand supersonically into the ambient IGM 
(``overpressured cocoon": \citealp{blandford1974,begelman1989, nath1995, heinz1998, yamada1999}). 
It is expected to form a shell of IGM matter compressed by the external shock
exterior to the radio lobes.

Detections of inverse Compton scattered photons in the X-ray energy bands of radio galaxies 
can drastically improve our understanding of the properties of hotspots and radio lobes.  
The distribution of energy among the magnetic field component and in the non-thermal 
electrons has been investigated by examining the spectral energy distribution from the radio to 
X-ray bands (e.g., for Cygnus A,  \citealt{harris1994,wilson2000}; for 3C123, \citealt{hardcastle2001}; 
for 3C295, \citealt{harris2000}; for Pictor A, \citealt{wilson2001}; for 3C120,  \citealt{harris1999}; 
for 3C390.3, \citealt{harris1998}).  The results of these studies revealed that high energy electrons 
of $\gamma\approx 10^3 - 10^5$ have an energy comparable to or greater than that in the magnetic field 
\citep[e.g.,][]{tashiro1998, grandi2003, croston2005, isobe2006,isobe2009}.
If the energy spectrum of non-thermal electrons is given by a power-law ($dN(E)/dE \propto E^{-\eta}$), 
most of the non-thermal electron energy is carried by those electrons characterized by energies 
near the lower cut-off energy of $N(E)$ for $\eta>1$.
However, direct measurements of the thermal and/or low energy ($\gamma\approx 10^3$ or less) 
electrons are quite difficult, since the plasma density in a radio galaxy is too low to emit 
detectable radiation.
The standard overpressured cocoon models of FR-II radio galaxies are characterized by two types 
of shocks, i.e., the external and the internal (jet-terminal) shock, where the latter is believed 
to be observed as hotspots. 
In this model, radio galaxies emit synchrotron radiation by shock accelerated electrons with 
an acceleration efficiency $\xi_e\equiv U_\mathrm{syn}/U_\mathrm{tot}$, which is 
inferred to be low ($U_\mathrm{syn}$ and $U_\mathrm{tot}$ are 
energy of synchrotron-emitting electrons and total internal energy, respectively).
If we adopt the standard model and assume the low acceleration 
efficiency in both shocks (low $\xi_e$), we can expect that a large amount of thermal electron 
energy, as well as non-thermal, synchrotron-emitting electron energy. 
In order to investigate the energetics of active galaxies, it is quite important 
to measure all of the energy in the electron component, including those that emitting strongly. 

In this paper we employ the Sunyaev-Zel'dovich effect (SZE) as a tool to measure the energy of the 
electrons in a radio galaxy.  The SZE represents the spectral deformation of the Cosmic Microwave 
Background (CMB) radiation due to the inverse Compton scattering of these photons by the energetic 
electrons \citep{zeldovich1969} in the galaxy.  The intensity change of the thermal SZE is classically 
described as follows \citep{zeldovich1969}; 
\begin{eqnarray}
  \frac{\Delta I_{x}}{I_{x}} &= &\frac{xe^x}{(e^x-1)}
   \left[ x\left(\frac{e^x+1}{e^x-1}\right)-4\right]  y,  \label{eq:SZ} \\
  y &=& \int \frac{k_BT_e}{m_ec^2}\sigma_Tn_e dl \propto\int p_edl,  \label{eq:press}
\end{eqnarray}
where $x\equiv h\nu/k_BT_r$ is the non-dimensional frequency, $T_r$ is the temperature of the CMB, 
$k_B$ is the Boltzmann constant, $T_e$ is the electron kinetic temperature, 
$\sigma_T$ is the cross section of Thomson scattering, $n_e$ is the electron number density, 
$m_e$ is the electron mass, and $p_e$ is the electron thermal pressure, respectively, 
integrated along the line of sight.
Equation (\ref{eq:press}) shows that the Compton parameter $y$ is proportional to the 
sum of the thermal pressure of electrons along the line of sight.
Therefore we can estimate the thermal energy of the electrons contained in a radio galaxy 
with the thermal SZE.
Similarly, the decrease in CMB intensity by SZE in the Rayleigh-Jeans regime reflects all of the electron
energy, not only the high energy electrons which generate X-ray photons, 
but also the lower energy electrons and the thermal electrons.
Although the energy distribution of the electrons is lost in the SZE (Eq.[\ref{eq:press}]),  it 
measures the total energy deposited in the radio galaxy.
In this paper, we ignore the non-thermal and kinetic SZE as well as relativistic corrections and focus 
mainly on thermal SZE as the first trial observation \citep{birkinshaw1999,ensslin2000,yamada2001}.

The study of the SZE has been directed toward understanding the thermal properties of the intra 
cluster medium (ICM), and it has been detected in dozens of clusters of galaxies (see for  recent reviews, \citealt{birkinshaw1999}; \citealt{carlstrom2001}; \citealt{rephaeli2002}).
Among the numerous SZE detections in clusters of galaxies, \citet{mckinnon91} first tried to detect the 
SZE in the radio lobes with the NRAO 12-m telescope at 90 GHz using the double-subtraction method.
They observed 4 FR-II radio galaxies whose lobe sizes were smaller than the beam size,  and obtained upper limits on the antenna temperature fluctuation ($\approx 0.1$ mK, or $y\approx 10^{-4}$). 
In this paper, we report on the results of the imaging observation of a giant radio galaxy B1358+305, 
extending much further in angular size than the beam size ($\sim 80^{\prime\prime}$ in our observation) using the Nobeyama 45-m telescope at 21 GHz. 
A two dimensional imaging study of a region in a radio galaxy is expected to provide  
the most reliable limit 
on the SZE in a radio galaxy by resolving the substructure within and the emission sources in the field of view.

The organization of the paper is as follows.
In Section 2 we briefly review the overpressured cocoon model \citep[see for detail,][]{yamada1999}.
In Section 3 we describe the features of our target B1358+305, and the observation and analysis procedures.
In Section 4 the results are presented.  We discuss the possible uncertainties in the estimation of the 
SZE amplitude in B1358+305 and their implications for radio galaxy models in section 5. Finally, we  
summarize our conclusions and future prospects in the last section.

\section{Overview of Overpressured Cocoon Model} \label{sect:model}

To provide an interpretative framework, 
we review our model of the overpressured cocoon \citep[see for details][]{yamada1999}, 
which is an extension of the classical overpressured-cocoon model of \citet{begelman1989}.
In this model, the jets from the AGNs form external shocks at the interaction with the surrounding IGM. 
Due to the action of the reverse shock (or internal shock), the jets are compressed.  In the FR-II radio 
galaxies, the compressed jet matter at the locations of the internal shock is believed to appear 
as hotspots.  The IGM compressed by the external shock is in pressure balance with the internal 
shock compressed jet matter at a contact discontinuity.  A high pressure gas clump forms around the 
AGN, which expands supersonically into the IGM \citep{begelman1989,nath1995,heinz1998,yamada1999}.
Hereafter we use the term ``cocoon" as the region consisting of both the jet-supplied matter and the 
shock-compressed IGM throughout this paper.\footnote{Note that the definition of the term ``cocoon" 
can change with different authors, and should not be confused with that defined herein.}
The cocoon expands as its internal energy, supplied by the jet, increases with time.
In this paper, for simplicity, we first treat $\xi_e$ is a single free parameter that characterizes the cocoon, 
and assume that the cocoon has an approximately spatially uniform thermal electron distribution. 
Thus the parameter $\xi_e$ is to be taken as an appropriately intensity-weighted mean value over the cocoon, and 
hereafter we denote the mean value as $\langle \xi_e\rangle$. 
This assumption may not be fully consistent with current observations, but 
its simplicity can provide a basis for feasibility evaluation and interpretation as a first 
step (see Section \ref{sect:alt} for the discussion of this assumption).

In these models, most of the matter that has passed through the shock is assumed to be 
almost fully thermalized except for a small fraction, which is assumed to be non-thermal. 
The density in the cocoon is insufficient for 
effective radiative cooling and, hence, the cocoon is expected to conserve nearly all of the 
kinetic energy of the jets. Furthermore, the cocoon is expected to remain hot for a relatively 
long time due to the low radiative cooling efficiency (an effective life time could be $t_\mathrm{life}
\sim 10^8$ yrs or more: see \citealp{yamada1999}).  Even taking into account the $PdV$ work 
against the surrounding IGM, the cocoon would remain hot for $\approx 10^8$ yrs \citep{yamada1999}.
Therefore the energy contained in a cocoon can be written as
  \begin{equation}
  U_\mathrm{tot} \approx L_\mathrm{jet}t_\mathrm{life},  \label{eq:etot}
  \end{equation}
where $L_\mathrm{jet}$ is the kinetic luminosity of the jets, and $t_\mathrm{life}$ is 
life time scale of the AGN (typically $\sim 10^7 - 10^8$ yrs).

For the thermal SZE as written (\ref{eq:press}) with the thermal pressure in the cocoon and the 
low value of $\langle\xi_e\rangle$, the total energy of the cocoon becomes $U_\mathrm{tot}= 
U_\mathrm{th}/(1-\langle\xi_e\rangle)=\int p_edV/(1-\langle\xi_e\rangle)$
which is proportional to $y$ (Eq.[\ref{eq:press}]).
This estimate indicates that SZE, which is proportional to $\int p_edl$, provides a 
good measure of thermalized electrons and the jet-supplied energy contained in the cocoon.
We employ this interpretative framework and attempt to measure the energy contained in electrons 
in a radio galaxy.

Since direct measurement of $L_\mathrm{jet}$ from observation is difficult,  equation (\ref{eq:etot}) 
is not used to estimate the expected value for the Compton $y$ parameter. Instead, we use the 
synchrotron emission argument and make the assumption that the synchrotron-emitting electrons are 
generated by diffuse shock acceleration with a small acceleration efficiency $\xi_e$ (again, we 
assume $\xi_e$ is the same for the internal and external shocks).  The minimum energy of the 
synchrotron-emitting electrons $U_\mathrm{syn}$ can be estimated from the synchrotron luminosity 
$L_{\rm syn}$ using the minimum energy condition, which in the simplest form \citep{moffet1975} is 
given by
\begin{equation}
 U_\mathrm{syn} = \frac{1}{2} 
 \left\{ \frac{9a}{2} \left( \frac{\pi c}{3e^7} \right)^{1/2} m_ec^2
   \frac{\alpha +1}{2\alpha +1} 
   \left( \frac{\nu_u^{\alpha+1/2}-\nu_l^{\alpha+1/2}}{\nu_u^{\alpha+1}-\nu_l^{\alpha+1}} \right)
 \right\}^{4/7} V^{3/7} L_\mathrm{syn}^{4/7}, 
\end{equation}
where $a$ is the ratio of energies of non-thermal protons and non-thermal electrons, 
$\alpha$ is the photon index of synchrotron emission, $\nu_u$ and $\nu_l$ are 
the upper and the lower limits of observation frequencies, and $V$ is the volume of the 
synchrotron-emitting region, respectively.
Once we obtain the minimum energy $U_\mathrm{syn}$, we can estimate 
$U_\mathrm{tot}\gtrsim U_\mathrm{syn}/\langle\xi_e\rangle$, where the inequality takes into account the larger
internal energy compared with the minimum energy.
If we adopt the small acceleration efficiency of the shock-accelerated electrons $\langle\xi_e\rangle$, 
the thermal electron energy contained in a cocoon is evaluated by Compton $y$ parameter as, 
 \begin{eqnarray}
   y &=& \frac{\sigma_T}{m_ec^2}\int p_e dl,   \nonumber \\
   &\simeq& \frac{\sigma_T}{m_ec^2} \frac{U_{\rm th}}{V} l
     = \frac{\sigma_T}{m_ec^2}(1-\langle\xi_e\rangle)\frac{U_{\rm tot}}{V} l
     \gtrsim \frac{\sigma_T}{m_ec^2} \frac{U_{\rm syn}}{V}\frac{1-\langle\xi_e\rangle}{\langle\xi_e\rangle} l, \\
   &\sim& 9.2\times 10^{-5} \left( \frac{l}{{\rm 300kpc}}\right) 
                                         \left( \frac{\langle\xi_e\rangle}{0.05}\right)^{-1} \left(\frac{U_\mathrm{syn}/V}{9.7\times 10^{-14}\, \mathrm{ergs ~ cm}^{-3}}\right), 
                                         \label{eq:expect}
 \end{eqnarray}
along the line of the sight that passes through the center of the cocoon.
Numerical values are based on the assumption that the axial ratio of the cocoon is 1:3 
(Figure \ref{fig:all}) and the adopted cosmological parameters ($H_0=72$  km  s$^{-1}$ Mpc$^{-1}$, 
$\Omega_M = 0.3$, $\Omega_\Lambda = 0.7$, \citealt{spergel2003}).
For the mean acceleration efficiency $\langle\xi_e\rangle$, we take the small value that is observationally suggested  
($\langle\xi_e\rangle \sim 0.05 - 0.1$: \citealp[e.g.,][]{sturner1997,baring1999,loeb2000}). 
The synchrotron-emitting electron energy in B1358+305 was estimated to be $U_\mathrm{syn}/V\sim 5.7-9.7
\times 10^{-14}$ ergs cm$^{-3}$ in the 10 MHz $\le \nu\le$ 10 GHz band by \citet{parma1996} 
(Yamada \& Parma, private communication),  
who assumed that the energy ratio of non-thermal protons and electrons, and the filling factor 
of the synchrotron electrons are equal to be unity.
Note that equation (\ref{eq:expect}) indicates the Comptonization parameter $y$, or 
the amplitude of the intensity decrement of SZE is proportional to the length within the 
radio galaxy along the line of sight, so that a giant source can yield a large value of $y$. 

\section{Observations and Data Reductions}

\subsection{Strategy of Search for SZE in a Radio Galaxy}
Our basic plan of observation in search for the SZE signal in a radio galaxy and the feasibility 
evaluation are provided below.  Compared to the case of galaxy clusters, the spatial extent of the 
cocoon relative to the radio lobes is less understood both theoretically and observationally 
\citep[see][and references therein]{mcnamara2007}.  We, therefore, develop the basic observational 
plan paying a special attention to the extent of the cocoon as follows.

\noindent
1) The amplitude of the intensity decrement of SZE is proportional to the length along the line of 
sight (Eq.[\ref{eq:expect}]).  Hence a large radio galaxy should be selected as a target. 

\noindent
2) In order to reduce the cancellation of the intensity decrement of SZE by the emission from the radio 
lobes, the observation frequency should be higher than that used in the observations of the synchrotron 
emission of the radio lobes.  Synchrotron emission has a power-law spectral energy distribution $\propto 
\nu^{-\alpha}$ with a typical $\alpha$ about 0.7, and its intensity decreases with $\nu$ in the GHz regime.

\noindent
3) Taking into account the potentially wider spatial extent of the cocoon in comparison to the radio lobes
that is assumed in the classical hydrodynamic models \citep[e.g.,][]{begelman1989}, a wider region than the 
radio lobes should be mapped.

\noindent
4) Since the expected signal of the SZE is weak and the cocoon structure is not well understood, 
substructures and emission sources within the mapped region should be carefully resolved and removed. 
To resolve the substructures and to take account of the proportionality of the $y$ parameter to the 
length scale along the line of sight (Eq.[\ref{eq:expect}]), a giant radio galaxy whose angular size 
is also large should be selected as a target.

In addition to the point 2, the amplitude of the thermal SZE is known to be maximum at $\nu\simeq 90$ GHz
\citep{birkinshaw1999}, hence the first choice for the observation frequency is 90 GHz.  However, the beam 
size decreases with the observation wavelength, and thus the observation frequency cannot be too high in 
order to perform the imaging observation of a wide region (the point 4) in a reasonable timescale.
The beam size at 90 GHz of the NRO 45-m telescope is about 20$^{\prime\prime}$, which is too small to 
image the large central region of B1358+305 ($\gtrsim 5^{\prime}$) to achieve a reasonable signal-to-noise 
ratio within a realistic timescale (see discussion on the feasibility in Section \ref{sect:feasibility} below). 
Additionally the atmospheric stability becomes worse for higher frequencies in 
this waveband, or the typical system temperature $T_\mathrm{sys}\approx$ 250 - 900 K
at 80 - 90 GHz is much higher than that at 21GHz ($T_\mathrm{sys}\approx$ 100 - 150 K). 
Hence, the required observational time becomes unrealistically long. 

We chose the single-dish observation as the first trial rather than interferometric observation.
This follows from the fact that the hydrodynamic models predict a smoother distribution of SZE 
compared to galaxy clusters \citep{begelman1989,yamada1999}.  In other words, the smooth SZE signal 
would be resolved out if an interferometric observation were carried out.  However, interferometric 
observations would be useful for detecting the SZE for more compact objects with known structures.
In addition, high angular resolution observation with interferometers can also be used to remove 
emission source contaminations within a field of view. Both single-dish and interferometric observations 
have their advantages and disadvantages.  In order to avoid the risk of resolving out of the possibly 
extended SZE signal, we use the single-dish observation strategy to measure the total flux.

Given these considerations and the low luminosity at 10 GHz, we decided to observe the central region 
of a giant radio galaxy B1358+305 with the NRO 45-m telescope at 21GHz.  The typical beam size of the 45-m 
telescope is $\sim 80^{\prime\prime}$ at this frequency.  The low system temperature (typically $\sim$ 
100 - 150 K) and a wide bandwidth (2 GHz) of the HEMT 22 receiver installed on the 45-m telescope led 
to a decrease in the necessary observation time for an SZE search in a faint giant radio galaxy.
The detailed set-ups and observation method are described in Section \ref{sect:obs} below.

\subsubsection{Feasibility Estimation} \label{sect:feasibility}
The imaging sensitivity for the observation with the 45-m telescope at 21 GHz is about  
$2T_\mathrm{sys}/\eta/\sqrt{B} =$ 7.55 mK $\mathrm{s}^{1/2}$, where the typical system 
temperature $T_\mathrm{sys}$ is taken to be 135 K, the 
aperture efficiency $\eta$ is estimated to be 0.8, and the bandwidth $B$ 
is taken to be 2 GHz for one of the dual channels, respectively. 
In order to obtain a 3$\sigma$ detection of the expected intensity fluctuation from equation 
(\ref{eq:expect}) or 0.30 mK with the beam width $\Delta\theta_\mathrm{HPBW}\simeq 
80^{\prime\prime}$ in the central $6.7^{\prime}\times 6.7^{\prime}$ region, the 
imaging sensitivity estimated above requires an observation time of about 67.3 hrs. 

The estimated observation time implies a deep imaging. 
As the expected SZE decrement in the B1358+305 is small, we examine the cumulative 
contributions from background (or foreground) discrete sources to the intensity fluctuation.
\citet{franceschini89} calculated the expected fluctuations of the antenna temperature 
($\Delta T_A/T_A$) as a function of beamwidth at 6\,cm, produced by randomly distributed 
sources within a flux range $10^{-8}<S({\rm Jy})<10$ (see Figure\,\,3 in \citealt{franceschini89}).
We assume that contributing sources have a spectral index $\alpha=0.7$ on average ($I_{\nu}\propto 
\nu^{-\alpha}$), and estimate the fluctuation of the brightness temperature $\Delta T_B/T_B (\sim \Delta T_A/T_A)$ 
at 21 GHz to be $1.1\times 10^{-4}$.
In terms of Compton $y$ parameter, this temperature fluctuation corresponds to $y=4.7\times 10^{-5}$ from equation (\ref{eq:SZ}).
The evaluated confusion limit is only marginally smaller than the expected 
SZE (Eq.[\ref{eq:expect}]). 
Hence in the data analysis procedure, emission sources in the field of view should be 
carefully removed from the integration.

\subsection{The Giant Radio Galaxy B1358+305}
B1358+305 is one of the largest radio galaxies of FR-II type at a  
moderate redshift ($z=0.206$, \citealt{parma1996}; \citealt{saripalli1996}).
Employing cosmological parameters for the $\Lambda$-dominated flat universe \citep{spergel2003}, 
its projected size amounts to $\sim$ 926 kpc, extending over $\sim 10^{\prime}$ in the plane 
of the sky \citep{parma1996}.\footnote{\citet{parma1996} estimated the size of B1358+305 to be 1.34 Mpc making use of cosmological parameters $H_0=100$ km s$^{-1}$ Mpc$^{-1}$ and $q_0=1.0$. 
We update cosmological parameters in this paper.} 
It has a relatively low radio luminosity ($P_{\rm 1.4GHz}=1.9\times 10^{25}$W Hz$^{-1}$), 
and the spectral age was estimated to about $\sim$2.5 -- $7.5\times10^7$ yrs from the multi-band imaging 
observations \citep{parma1996}.
\citet{parma1996} further calculated the advance speed of the shock head of 
B1358+305 as $R/\tau_\mathrm{spec}\simeq 0.02c$ -- $0.03c$ ($R$ is the lobe length, $\tau_\mathrm{spec}$
is the spectral age, $c$ is the speed of light, respectively). 
From the balance of the external ram pressure and the jet ram pressure, they concluded that 
the environment density of B1358+305 is quite low ($n_e\lesssim 10^{-6}$ cm$^{-3}$), and 
B1358+305 is overpressured even for transverse direction, for an ambient temperature as 
high as that of a typical poor cluster environment ($\sim 10^7$ K).
This makes B1358+305 a good target in the sense that it resembles the classical model.

B1358+305 has an FR-II morphology with two aligned lobes with a radio core, 
which resides closer to the northern lobe.
As SZE appears as the decrement of the CMB in the frequency range $\nu\lesssim$ 200 GHz, 
any compensating radio emission should be avoided.
Since the lobe intensity rapidly decreases towards the region slightly south of the core,  
we focused on that region in order to avoid the lobe emission contamination as much as possible
(Figure\,\ref{fig:all}).

\subsection{Observations} \label{sect:obs}
Observations of B1358+305 were performed with the HEMT 22 receiver on the Nobeyama 45-m telescope from 2001 
February 28 to March 18.  The receiver has 2 GHz bandwidth, and is equipped with dual channels which can 
simultaneously receive two  circular polarization components, which enables the reduction of the required 
observation time.  The near-central region in B1358+305 of size $6.7^{\prime}\times 6.7^{\prime}$ around 
($\alpha_{50}, \delta_{50})=(13^{\rm h}58^{\rm m}25.0^{\rm s}, +30\degr 32^{\prime}0.0^{\prime\prime}$) was 
raster-scanned with $40^{\prime\prime}$ spacing, which created a map composed of $11\times 11$ pixels.
Since the jet axis is nearly along the declination, raster-scans were performed along the RA-DEC coordinates in turn.
The scan speed is about $25^{\prime\prime}$ s$^{-1}$, resulting in the acquisition of one map in $\sim 180$ s. 
The exposure time was about 74.8 ksec in total, corresponding to 618 s for each pixel.  The system noise 
temperature was about $T_{\rm sys}=140$\,K on average, ranging from about just below 130 K to nearly 150 K.
The noise produced by the fluctuations in the atmosphere were reduced by the simultaneous observation of an 
off-source point 
via the beam switching technique.  The off-source point 
is about 400$^{\prime\prime}$ away in azimuth direction to the east and was simultaneously observed at 15 Hz. 
The antenna temperature was calibrated using the chopper-wheel method, while the optical depths of the 
atmosphere were measured by elevation scans.  Since the atmospheric optical depths were about 0.04 throughout 
the observing period,  the flux was not corrected for atmospheric absorption.
The main beam size $\theta_{\rm HPBW}$ was estimated to be $81.2^{\prime\prime}$ by observing 3C273.
With this beam size we can image a large ($6.7^{\prime}\times 6.7^{\prime}$) area in a 
reasonable timescale, simultaneously resolving the substructures in the mapped region.
For the pointing and flux calibrator, we used 3C286 about $7.5\degr$ away from the center of the image.
The pointing offset was monitored about every hour, and  was typically smaller than $6^{\prime\prime}$ 
throughout the observing period.  The flux of 3C286 was assumed to be $2.56\pm 0.2$Jy \citep{ott1994}, 
and it was used to evaluate the aperture efficiency of the telescope.
Due to the good weather conditions, the aperture efficiency was stable during the latter half 
of the observation period (rms of the flux variation is about 5\%).
In the former period, the stability was slightly worse (the flux variation reached about 10\%), and we 
carefully corrected each observation using the efficiency measured just before the observation.

\subsection{Data Reduction and Analysis}  \label{sect:analys}
Raster-scans were performed in two orthogonal directions (along the right ascension and the declination coordinates) 
in turn in order to obtain pairs of orthogonally scanned maps taken as close together in time as possible. 
Within a raster-scanned map, antenna temperature fluctuations induced by the time variation of the atmospheric 
conditions, instability of the detectors or other instruments and so on, appeared along the scanning direction 
(scanning noise). The scanning noise in the data was reduced by combining the two orthogonally scanned maps
in the Fourier space via the basket-weaving method (\citealt{sieber1979}; \citealt{kuno1993}).

Although we tried to avoid contamination from lobe emissions into the mapped region, some of the scans suffered 
from lobe emissions in the operation of beam switching.  Due to the limited throw amplitude ($\sim 400^{\prime\prime}$) 
and to the rotation of the position angle of the beam switch, emissions from one of the lobes entered into the 
reference field, resulting in over-subtraction in a part of the mapped region (less than 1/4 of the mapped area at 
most).  This over-subtraction generated artificial intensity decrements. 
We identified the affected pixels in each scan and removed them in the integration.
In the identification process, we made use of the 10.6 GHz map of \citet{saripalli1996} 
(see Figure\,\ref{fig:all}), and set the position of the reference beam $400^{\prime\prime}$ away from the main beam.
Since at 21 GHz both lobes are expected to be dimmer than displayed in the map in 
\citet{saripalli1996} (the spectrum is a decreasing function of  frequency), 
the procedure above does not overlook the affected pixels.
Due to this partial masking in the integration, the noise level is not uniform in the 
obtained integrated map, ranging from 0.72 mJy beam$^{-1}$ to 1.2 mJy beam$^{-1}$ 
(1$\sigma$ level): in spite of the loss of some pixels, however, the root mean square of the 
intensity in the integrated map reached 1.22 mJy beam$^{-1}$ with emission sources.

In order to evaluate the noise level without emission sources, 
we constructed two independent maps by dividing the maps taken in the former 
and the latter halves of the observing period, and subtracted the latter from the former.
In this process, emission from true sources should cancel out and
the resultant map should contain only random noise.
The rms fluctuation of the differential map constructed in this way is 1.01 mJy beam$^{-1}$.

\section{Results}
Figure \ref{fig:full_image} shows the integrated map of the central region of B1358+305.
One can easily identify the AGN component and the innermost edges of the two radio lobes 
in this figure, which appear also in Figure \ref{fig:all}.
On the right edge of the map exists a bright region, with a peak intensity of 
$\sim 6$ mJy beam$^{-1}$ (source ``D"). 
We failed to find any identified source corresponding to this bright spot in any 
catalogue from the radio to X-ray energy band, though, it indeed appears on the VLA NVSS 1.4 GHz 
map and the Effelsberg 10.6 GHz map obtained by \citet{saripalli1996}.
Taking into account a relatively large beam size ($\theta_{\rm HPBW}=81.2^{\prime\prime}$), 
we recognize it as an unknown point source.

\begin{figure}
    \plotone{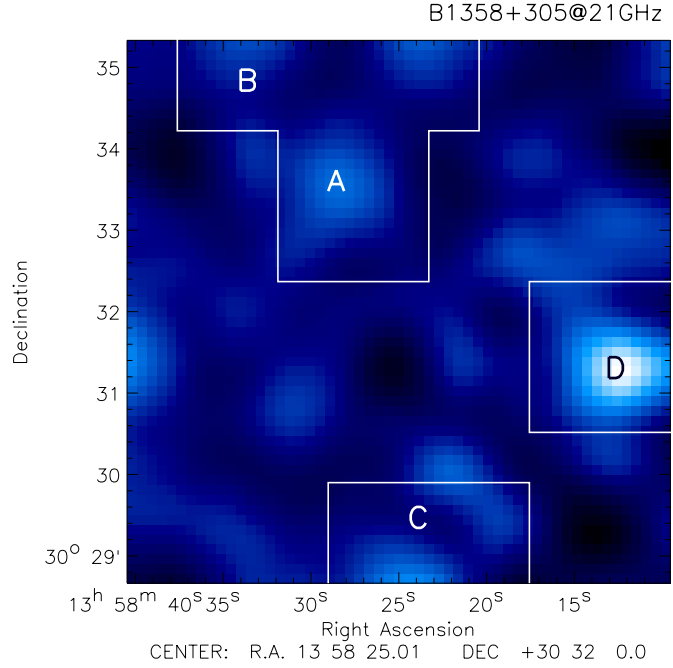}
  \caption{Locations of the emission sources. Masked region corresponding to the AGN, 
  north and south lobe edges, and the unknown source on the western side (labeled ``D" in the integrated map)
  are indicated with solid white lines.}
  \label{fig:mask}
\end{figure}

Since emissions from the sources smear out the SZ flux decrement, it is necessary to subtract 
them from the map.
However, it is quite difficult to obtain the correct zero flux level for mainly two reasons.
One is related to the larger uncertainties in models of cocoons in comparison with  clusters of galaxies.
In the case of a galaxy cluster, a well-established model obtained by X-ray observation (so-called $\beta$ model) 
is available.  Thus, one can set the baseline (zero flux level) at a sufficient distance ($\theta\gg \theta_{\rm core}$;
see e.g., \citealt{komatsu2001}) from the center with the help of the X-ray observation. 
On the other hand, in the case of the theoretical model for the cocoon, there are many uncertainties in its 
morphology.  For example,  the density distribution of the surrounding 
medium \citep[e.g.,][]{kaiser1999} or the magnetic fields alter the shape of the cocoon 
(see \citealt{burns1991} for a review of numerical simulations, and for examples of MHD simulations, see e.g.,  \citealt{clarke1986}; \citealt{lind1989}). 
The other reason is that our deep imaging of this region approaches the confusion limit, so the 
background fluctuation impedes the establishment of a definitive zero flux level (see Section \ref{sect:feasibility}).
Therefore, we could not ascertain the absolute intensity of the sources included in the map at 21 GHz, and 
it is difficult to correctly subtract the emission from the true emission sources.
Instead we evaluated the intensity fluctuation in the mapped region, and re-defined 
the baseline so that the average fluctuation became zero. 
In order to calculate the intensity fluctuation, we masked the emission source regions. 
These regions around the emission sources are indicated by white lines in Figure \ref{fig:mask}.
The baseline of the map is re-defined after the source removal so as to set the average 
intensity of the remaining pixels to be zero.

\subsection{Structure of the Cocoon}  \label{sect:structure}
\begin{figure}
   \epsscale{.60}
    \plotone{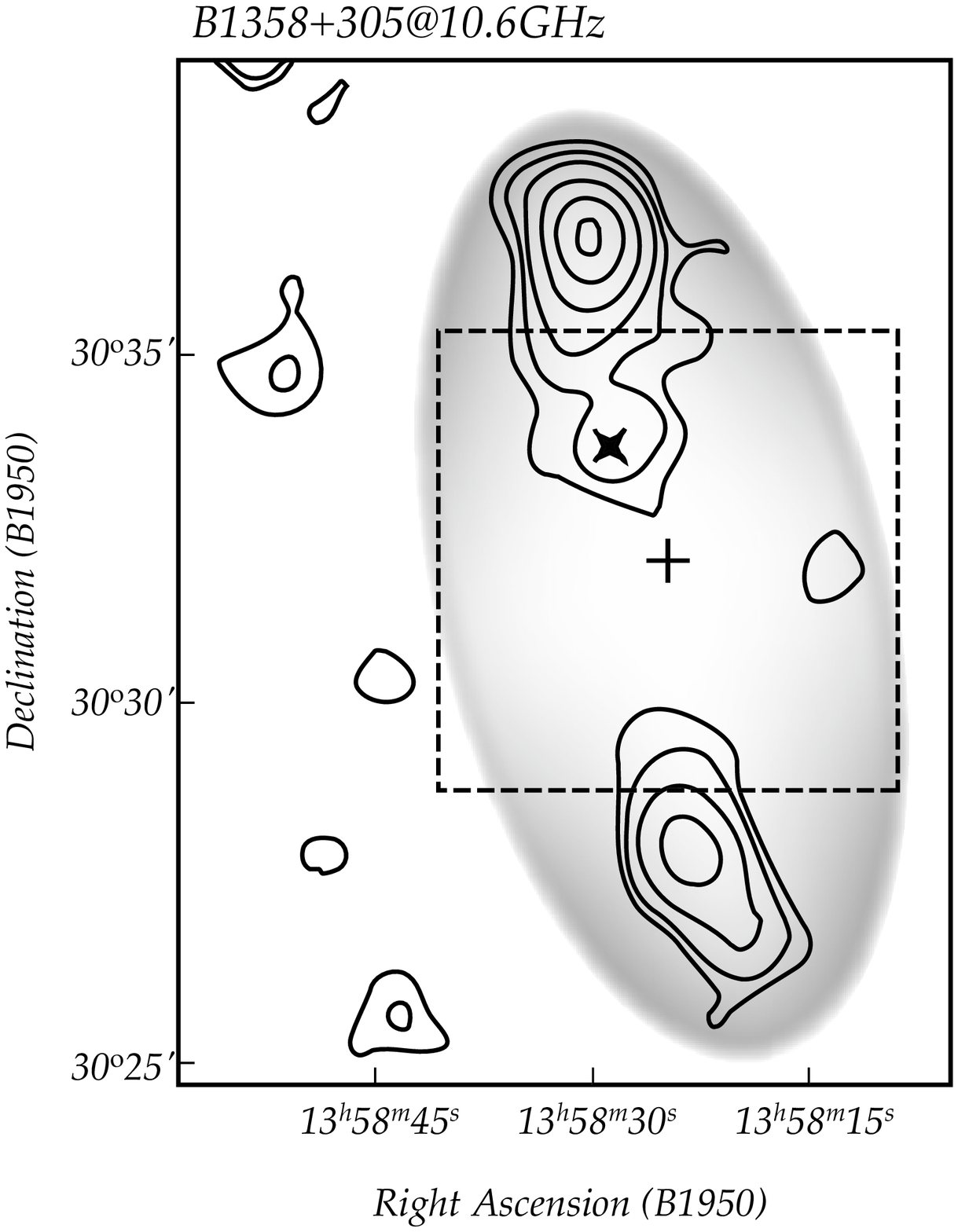}
  \caption{A full image of B1358+305 observed with the Effelsberg 100-m telescope at 10.6 GHz (solid contour: reconstructed from an original image in \citealt{saripalli1996}).
 The region we observed ($6.7^{\prime}\times 6.7^{\prime}$) is indicated by the dashed-line square 
 around the center (indicated by a cross) at ($\alpha_{50}, \delta_{50})=(13^{\rm h}58^{\rm m}25.0^{\rm s}, +30^{\circ}32^{\prime}0.0^{\prime\prime}$). 
 A fat cross in the northern lobe indicates the position of the radio core. 
 A schematic picture of the cocoon is overlaid as a gray elliptical figure.}
  \label{fig:all}
\end{figure}

In order to improve the effective signal-to-noise ratio, we constructed projections of the 
two-dimensional map along two orthogonal directions.
We can also expect that these projections reflect the structure of the cocoon more clearly 
than the original two-dimensional map if the data contains the SZE signal from the cocoon of B1358+305.
As can be seen in Figure \ref{fig:all}, the jet axis is almost perpendicular to the right ascension.
Thus, we can examine the structure perpendicular (parallel) to the jet axis by projecting the map along 
the declination (right ascension) coordinates.  In the process of projection, we average the pixels along 
each column or row without using the masked pixels corresponding to the emission sources that are indicated 
with the white lines in Figure \ref{fig:mask}.

\begin{figure}
    \plotone{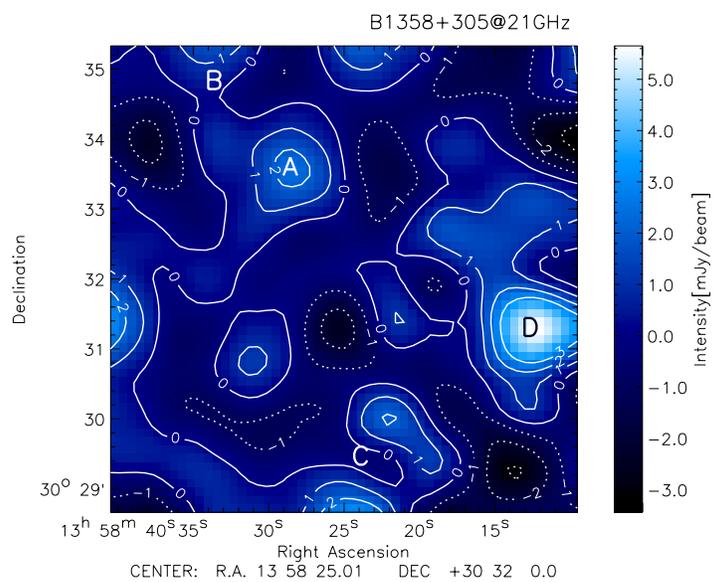}
  \caption{Integrated map of the central region of B1358+305 at 21 GHz.
  Labels A, B, C and D indicate the AGN, north and south lobe edges, and the unknown source (see text), respectively. 
Contours are $-3,-2,-1,0,1,2,3$ mJy beam$^{-1}$ (see numbers on the contours). 
Intensity is normalized so that the average is zero in the observed region, and the contour levels relative intensities.}
  \label{fig:full_image}
\end{figure}

\begin{figure}
  \plotone{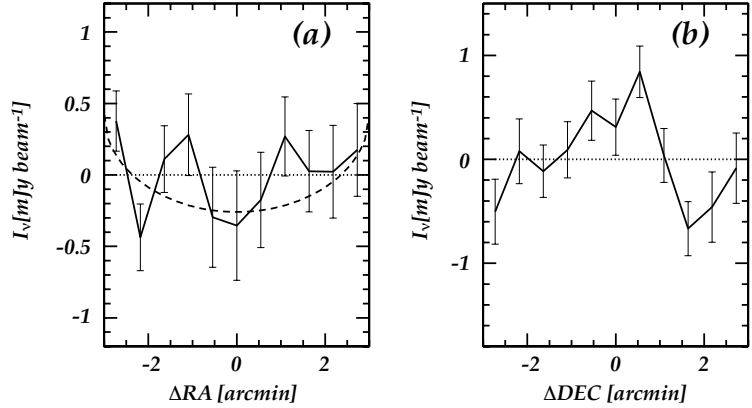}
  \caption{Projections along (panel (a)) and onto (panel (b)) the jet axis. Error bar is 1$\sigma$. 
  In panel (a), relative decrement at about $+2^{\prime}$ west of the central column is mainly due to the dark region north of the source labeled ``D" and 
  west of the southern lobe edge (labeled ``C"). A dashed curve denotes an expected decrement for a 
  cylindrical isobaric cocoon, and a dotted line is the flat intensity (no SZE decrement).
  In panel (b), there is a peak about $0.7^{\prime}$ north of the central row and two decrements 
  toward both edges. These decrements are due to two dark regions north and south of the 
  unidentified source ``D" (see Figure \ref{fig:full_image}).}
  \label{fig:proj}
\end{figure}

Figure \ref{fig:proj} (a) shows the projection along the jet axis (the declination coordinate).
A relatively dark region appears at about $+2^{\prime}$ west of the central column. 
Comparing Figure \ref{fig:proj} (a) with the original two-dimensional map (Figure\,\ref{fig:full_image}), 
one can see the decrement appearing in Figure \ref{fig:proj} (a) mainly picks up 
the dark region between the unidentified source labeled ``D" in the map and the AGN (labeled ``A").
However, as the statistical errors are large (error bars in Figure \ref{fig:proj} correspond to $1\sigma$) 
even though they are reduced by $\sim 1/\sqrt{N_p}$ in averaging over each column ($N_p$ is the number of 
pixels included), we cannot rule out the possibility that the projection obtained here is consistent 
with a flat distribution (no SZE signal accompanied B1358+305).
As shown in Figure \ref{fig:all}, B1358+305 has a relatively large axis ratio, and it is reasonable to 
approximate it as a cylinder with an axis along the jet axis.
Hydrodynamic simulations \citep[e.g.,][]{loken1992,kaiser1999,scheck2002} show that 
the pressure $p_e$ is nearly uniform inside a cocoon 
when the density distribution of the surrounding IGM is nearly flat
(there will be a difference in $\xi_e$ at internal and external shocks, and accordingly $p_e$ 
would not be spatially uniform as is suggested by observation, but this assumption is discussed 
separately in Section \ref{sect:alt} below).
Since $y\propto p_el$ (Eq.[\ref{eq:press}], where $l$ is the length cut through the cocoon along the line 
of sight), the projection along the jet axis is expected to present $l(\theta)$ in the isobaric cocoon, 
where $\theta$ is the angular distance projected onto the sky from the map center.
For a cylindrical cocoon, therefore, the intensity decrement $\Delta I_{\nu}$ is 
proportional to $-\sqrt{1-\theta^2}$.
We overplot the expected profile ($\Delta I_{\nu}\propto -\sqrt{1-\theta^2}$) based on equation (\ref{eq:expect}) 
with fiducial values on Figure \ref{fig:proj} (a) as a dashed line, offset so that the average intensity is 
equal to zero.
The obtained distribution is not inconsistent with the expected one or with the flat distribution.
This implies the dominance of the statistical errors over the signal reflecting the possible cocoon 
structure in the observed region.
We estimate the standard deviation from the flat distribution,
\begin{equation}
  |\Delta I_{\nu}| = 0.28 ~ {\rm mJy ~ beam}^{-1},  \label{eq:xsigma}
\end{equation}
which sets the upper limit $y_{\rm upp}\lesssim 1.04\times 10^{-4}$ at the 95\% confidence level
(Eq.[\ref{eq:SZ}]).

We further examine the allowed upper limit for a Compton $y$ parameter by a chi-square test
of the data for the projection along the jet axis and the expected profile from a cylindrical cocoon 
model $\Delta I_{\nu}\propto -\sqrt{1-\theta^2}$.
We find that as long as $\Delta I_\nu \le$ 0.32 mJy beam$^{-1}$ (or $y\le 1.19\times 10^{-4}$), 
these two distributions can agree at the 95\% confidence level. 
Herewith we denote these as $|\Delta I_\nu|^{\chi}$ and  $y_\mathrm{upp}^{\chi}$.

We have also constructed the projection onto the jet axis.
If the cocoon has a prolate figure and lies in the plane of the sky, 
the pressure scale length along the jet axis is larger than the declination range of 
the observed field (see Figure\,\ref{fig:all}). Hence, the projection onto the jet axis would display 
a more moderate decrement in contrast with the projection along 
the jet axis.\footnote{If the jet axis is not straight or the inclination angle is not small, 
the projected profile will be different from the one described here.
However, \citet{parma1996} concluded that a bend of the jet or the inclination are not so significant, from the contrast of the intensity of northern and southern lobes.}
Figure \ref{fig:proj} (b) shows the projection onto the jet axis.
There clearly exist a peak about $0.7^{\prime}$ north of the central row and two decrements toward both edges, 
which are not consistent with the expectation from the isobaric cocoon model filled with thermal electrons (Section 2).
These decrements are due to the two dark regions north and south of the unidentified source ``D" (at 
around $\alpha_{50}\sim 13^{\rm h} 58^{\rm m} 20^{\rm s}$: see Figure\,\ref{fig:full_image}).
The standard deviation from a flat distribution derived from the projection in this direction (onto the jet axis) is $|\Delta I_{\nu}| =$0.44 mJy beam$^{-1}$, which is larger than the statistical error estimated from the noise level of the original two-dimensional map in the process of the projection ($1.01/\sqrt{N_p}$, $\sim 0.27$\,mJy beam$^{-1}$ on average).
Even though the dark regions appearing in Figure \ref{fig:proj} may indeed suggest the existence of SZE  
signals, we cannot rule out the possibility that these are not due to B1358+305.

\begin{figure}
  \plotone{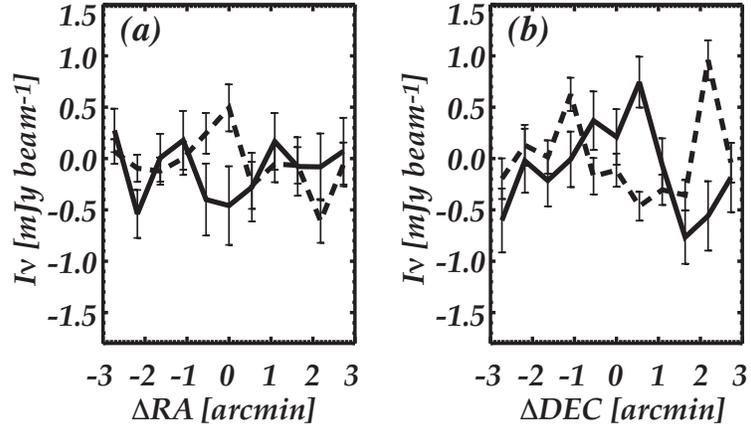}
  \caption{Projections along (panel (a)) and onto (panel (b)) the jet axis of the differential 
  map (dashed lines; Section \ref{sect:analys}). Solid lines are the same as Figure\,\ref{fig:proj}. 
  The differential map projections which consist of only random noises have similar amplitudes
  with the total map, but different distribution from it. This indicates the intensity fluctuations 
  are likely to come from the excess atmospheric noises (see text). }
  \label{fig:proj_diff}
\end{figure}

In order to examine the dominant source of the fluctuations in Figures\,\ref{fig:proj} (a) and 
\ref{fig:proj} (b), we analyzed the differential map in Section \ref{sect:analys} in the same 
manner as the total map. Specifically,  the differential map was constructed by dividing the observation period 
into the former half and the latter half. 
Figure\,\ref{fig:proj_diff} shows the fluctuations in the projections of the differential map 
overplotted with dashed lines on the total map (solid lines, the same as Figure\,\ref{fig:proj}). 
The projections on the both axes have similar amplitudes (the rms of the 
projection along the jet axis is 0.27 mJy beam$^{-1}$ and the projection 
onto the jet axis is 0.43 mJy beam$^{-1}$) as those of the total map, but 
their distributions do not resemble the total map. 
Taking into account the fact that real signals, including foreground or background 
emission sources in the field of view, do not appear in the differential map, 
we conclude that the fluctuations in Figure\,\ref{fig:proj_diff} are largely caused by the 
excess atmospheric noises at 21 GHz.

\section{Discussion}\label{sect:discussion}

Due to the lack of definite information about the position of the edge of the cocoon, it is 
difficult to define a concrete baseline.  For this reason we 
evaluated the relative variation of the intensity in the map.  In order to improve the 
signal-to-noise ratio, we made two projections on orthogonal directions.  A differential map 
analysis in Section \ref{sect:analys} shows that the main source of the relative fluctuations in 
these projections is the excess atmospheric noise rather than the true SZE signal induced by the 
hot electrons in B1358+305. 

We discuss possible factors that may reduce the true amplitude of SZE in B1358+305 in comparison 
to the estimate with equation (\ref{eq:expect}), including a non-uniform distribution of 
thermal electrons. 
In addition, the degree to which our results can affect the 
previous studies on pressure balance of B1358+305 (Section \ref{sect:pressure}) is examined.
Finally, we direct attention on radio galaxies in clusters of galaxies to demonstrate the usefulness 
of the SZE as a probe of the energetics of radio galaxy in a more general sense to guide 
future observations.  It is to be emphasized that the intensity fluctuations discussed in this 
section should not be confused with a true SZE signal in B1358+305, but only as a rough guide 
for a quantitative discussion.

\subsection{Thermal Evolution of the Cocoon}  \label{sect:cooling}
The expected amplitude of SZE is estimated using a steady mean acceleration efficiency $\langle\xi_e\rangle$ 
(Eq.[\ref{eq:expect}]).  
If $\langle\xi_e\rangle$ is significantly different from assumed, the expected value of $y$ is also affected. 
In particular, different cooling timescales for the thermal and non-thermal electrons may alter 
$\langle\xi_e\rangle$ from the fiducial value (=0.05 in Eq.[\ref{eq:expect}]) in time, especially in the central 
region where the oldest population is contained.  Since an underestimate of $\langle\xi_e\rangle$ leads to the 
overestimate of the expected $y$ amplitude (Eq.\,[\ref{eq:expect}]), we examine whether the 
different cooling timescales for the thermal and non-thermal electrons can significantly alter 
$\langle\xi_e\rangle$ using a simple argument about exhaustion of the internal energy.  Since the cooling 
timescale is strongly dependent on the electron energy $\gamma$, we discuss synchrotron electrons 
of high $\gamma$ ($\gamma\gtrsim 10^3$ or more) and thermal electrons ($\gamma\lesssim 10$ at 
$T_e\lesssim 100$ keV) separately. 

\citet{parma1996} derived the spectral ages of the non-thermal electrons for both the northern and 
southern lobes using the compiled multi-frequency observation of B1358+305 from 325 MHz to 10550 MHz, 
finding the break frequency to be 2.4 GHz from the integrated spectral energy distribution.  Further 
fitting of the multi-frequency images with the energy loss formula including synchrotron and 
CMB inverse Compton losses leads to spectral ages of $\tau_\mathrm{spec}$ for B1358+305 in the range 
$2.5\times 10^7$ yrs $\lesssim \tau_\mathrm{spec} \lesssim 7.5\times 10^7$ yrs. 
Since the synchrotron cooling timescale and inverse Compton cooling timescales are proportional to 
$\gamma^{-1}$, lower energy electrons have a longer cooling timescale.
The simple scaling of the above estimation by \citet{parma1996} indicates that an electron that 
emits 325 MHz synchrotron light has a spectral age of about a few times of $\sim 10^8$ yrs, or even in the 
extreme case $\sim 10^9$ yrs at best.  We adopt $10^8$ yrs as the typical cooling time scale for 
the non-thermal electrons. 

On the other hand, a relativistic Maxwellian distribution of high temperature thermal electrons,
\begin{equation}
  P_e(\gamma)d\gamma 
    = \frac{\gamma^5\beta^2\exp(-\gamma/\Theta)}{\Theta K_2(\Theta^{-1})}   
       d\gamma,  \label{eq:thermal}
\end{equation}
spans over $\gamma\lesssim 10$ for $k_BT_e\lesssim $100 keV (where $K_2(x)$ is the second order modified 
Bessel function, $\beta=v/c$, and $\Theta\equiv k_BT_e/m_ec^2$ is the non-dimensional electron temperature). 
We estimate the cooling timescales following \citet{sarazin1999}. 
Inverse Compton scattering of CMB photons, synchrotron self-Compton scattering, Coulomb cooling, and bremsstrahlung are taken into account as cooling processes.  The density of the IGM and in B1358+305 is quite low 
($n_e\lesssim 10^{-6}$ cm$^{-3}$; \citealt{parma1996}), and thus bremsstrahlung and 
line cooling are negligibly small compared with the other processes. 
In this low energy regime ($\gamma\lesssim 10$) and in the very poor environment (the density of the 
surrounding IGM was estimated to be $1.4\times 10^{-7}$ cm$^{-3} \lesssim n_\mathrm{IGM} \lesssim 
8.4\times 10^{-7}$ cm$^{-3}$; see Table 3 of \citealt{parma1996}), the major cooling mechanisms are 
inverse Compton cooling and the Coulomb cooling, but the cooling time is longer than the Hubble time 
($\tau_\mathrm{cool}\gtrsim 10^{10}$ yrs).
Therefore, the thermal energy of electrons is conserved for a typical cooling time of the 
synchrotron electron ($\sim 10^8$ yrs).
Even if we adopt a cooling timescale for electrons that radiate low frequency 325 MHz photons
($\tau_\mathrm{cool}\lesssim 10^9$ yrs), it is unlikely that the cooling time of thermal electrons 
is lower than that of non-thermal electrons.   
In other words, as long as the radio lobes of B1358+305 are bright in synchrotron emission, the 
thermal energy of electrons in B1358+305 does not change in time.
Therefore, it is unlikely that the lack of the detection of an intensity decrement in B1358+305 is 
attributable to the loss of thermal electrons by cooling.

In addition to the cooling processes described above, some models emphasize the importance of adiabatic 
expansion \citep{ito2008} or backflows of jet matter that escapes thermalization at the shock interaction 
\citep{scheck2002}.  These models predict a lower thermal pressure of electrons than the model of 
\citet{begelman1989}.  Quantitative evaluation of these effects has yet to be determined, but it is to be noted that our model 
tends to overestimate the thermal pressure of the electrons in a cocoon. 
If these effects that reduce the thermal pressure of electrons are significant  
in the evolution of a radio galaxy, much more sensitive observations will be necessary.

\subsection{Non-Thermal Electrons in Radio Lobes} \label{sect:alt}

In this paper, we adopted a model based on the classical overpressured cocoon model with 
a constant and uniform mean shock acceleration efficiency $\langle\xi_e\rangle$, and 
a uniform distribution of thermal electrons within a cocoon as the simplest and most 
optimistic interpretation of the observation (Section \ref{sect:model}). 
However, in spite of the search for X-rays in radio galaxies in galaxy clusters, 
the clear evidence for non-thermal electron acceleration at the external shock has not been 
found except for objects such as Cen A \citep{croston2009}.
Although this lack of clear evidence may reflect the limited observational sensitivity or angular resolution, 
the uncertainty in determining $\langle\xi_e\rangle$ is also likely to contribute. 
Especially, at the internal shock or termination shock of the relativistic jets, $\xi_e$ can be 
as large as close to unity, while $\xi_e$ at the non-relativistic external shock is much smaller.  
In this case, the radio lobes within the shocked shell would have only a small population of 
thermal electrons, whereas the shocked shell has a large fraction of thermal electrons as 
in the first adopted model (Section \ref{sect:model}). 

The spatial SZE profile of this model would differ significantly from that adopted (see Section 
\ref{sect:structure}).
The amplitude of SZE caused by non-thermal power law electrons depends on the parameters 
describing the electron energy distributions (e.g., the power law index, the lower and upper 
cut-off energies, and the break energy if any).
The ratio of thermal and non-thermal SZE amplitudes is consequently dependent on these parameters.
If the non-thermal SZE amplitude is much smaller than thermal SZE in the radio lobes, the SZE spatial profile for this case 
would not reflect the length passing through the cocoon, but the length passing through the shocked IGM shell. 
Therefore, it may have a flatter SZE distribution as compared to that in Section \ref{sect:structure} 
($\propto -\sqrt{1-\theta^2}$).  This possibility makes the detection of SZE much more difficult 
\citep[see e.g.,][]{pfrommer2005,colafrancesco2008,hardcastle2008}. 
On the contrary, it may also be possible that the non-thermal SZE amplitude is similar to the thermal SZE.
In this case, the spatial profile does not differ from that in Section \ref{sect:structure}, and detection of SZE is 
unlikely to be significantly more difficult compared with the cocoon model filled with thermal electrons (Section \ref{sect:model}).
Multi-frequency observation will be necessary to discriminate these distributions (see Section \ref{sect:xray} below). 
Additionally, it is possible that a mature giant radio galaxy does not expand supersonically 
in the lateral direction \citep[e.g.,][]{konar2009}.
In this case, the SZE signal peaks at the head of the external shock and decreases toward the central 
region where we observe (see below). 
Since the diffuse shock acceleration efficiency problem remains to be solved, we cannot 
exclude either of these scenarios with different acceleration efficiencies. 
In order to solve this problem and to obtain a deeper understanding of the subsequent dynamical 
and thermal evolution of a radio galaxy, progress in both theoretical studies of shock acceleration 
efficiency and higher sensitivity observations will be required. 

Radio images and some X-ray images of radio galaxies suggest that the central 
region close to the nucleus lack the energetic electrons that many hydrodynamic  
models suggest. 
If this applies to B1358+305, our results would correspond to genuine atmospheric noise. 
However, the X-ray images of a FR-II radio galaxy MS0735+74221 and Cygnus A 
show extended diffuse X-ray emissions that fill the central regions 
\citep[see for a review,][and references therein]{mcnamara2007}. In addition,  
\citet{hardcastle2008} observed bright FR-II galaxies at 90 GHz with the BIMA, and found 
that some of their sample (3C 20 and 3C 388) have extended emissions that cover the 
central regions. 
Although there are ambiguities in the distributions of the thermal and synchrotron electrons, 
these results imply that some FR-II radio galaxies seem to have electron populations 
at the central regions away from the jets and hotspots. 
We cannot tell if B1358+305 is deficient of electrons because of a small number of past observations 
or limited sensitivity of our results. 
Hence we conclude that our results are largely due to atmospheric noise, but we 
cannot completely rule out the small SZE signal in B1358+305, and much higher sensitivity 
is required in future observations.

\subsection{Expansion of B1358+305} \label{sect:pressure}

Since only upper limits on the $y$ parameter were obtained. we compare our results with past studies 
of B1358+305 in the case where (1) the true pressure is close to that derived from the upper limit, 
and (2) the true pressure is significantly lower than the obtained upper limit. 

Consider the first case.  Since the observed region does not extend over the entire cocoon, the 
perpendicular projection onto the jet axis will show a flatter intensity 
distribution than the parallel projection (i.e., the pressure scale length along the jet axis is 
longer than the declination range of the observed field: see Figure\,\ref{fig:all}).  Hence, it  
is more easily subject to the contamination from the background emission sources and the excess 
atmospheric noise.  On the other hand, the right ascension range of the observation is expected to be 
comparable to the cocoon width and the parallel projection along the jet axis is expected to represent 
the structure of the cocoon (Section 4.2).  Therefore, we employ the upper limit of the $y$ parameter 
calculated with the projection along the jet axis ($|\Delta I_\nu|_{2\sigma}=0.56$ mJy beam$^{-1}$, 
or $y_{\rm upp}\lesssim 1.04\times 10^{-4}$) in the discussion below.
If we assume that the obtained intensity fluctuation accompanies B1358+305, then 
the electron pressure derives $p_e = 2.20\times 10^{-12}$ dyne cm$^{-2}$ (Eq.[\ref{eq:expect}]) for 
$l$=300\,kpc. 
If we take values obtained by the chi-square test $y_\mathrm{upp}^{\chi}$, the equivalent electron 
pressure becomes $p_e = 2.51 \times 10^{-12}$ dyne cm$^{-2}$.  On the other hand, \citet{parma1996} 
deduced the jet ram-pressure to be $\ge 1.2\times 10^{-12}$ dyne cm$^{-2}$, which is comparable to the 
equipartition value under the assumption that B1358+305 expands supersonically. 
This result is quantitatively consistent with \citet{parma1996} in the sense that B1358+305 
expands supersonically against the surrounding IGM. 
However, the results obtained using the upper limit in our observation ($2.20-2.51\times 10^{-12}$ dyne cm$^{-2}$) 
are almost double the result of \citet{parma1996} ($\ge 1.2\times 10^{-12}$ dyne cm$^{-2}$).
This difference would come from the assumed parameter value $\langle\xi_e\rangle=0.05$ (Eq.[\ref{eq:expect}]), 
and may imply a larger mean acceleration efficiency ($\langle\xi_e\rangle \gtrsim 0.1$) in B1358+305 
and significant contribution from non-thermal SZE of smaller amplitude than thermal SZE.

Since the intensity fluctuations obtained turned out to be dominated by the excess atmospheric 
fluctuations, however, it is also possible that the true thermal electron pressure in B1358+305 
is much lower than that calculated with $y_\mathrm{upp}$ or $y_\mathrm{upp}^{\chi}$ (the case (2)). 
Because B1358+305 resides in a poor environment, it is difficult to directly measure the IGM pressure
around it.  Instead we estimate the external pressure under the assumption that the environment resembles  
an extremely poor galaxy cluster or a group of galaxies, and assume the IGM temperature to be 
$\lesssim 10^7$\,K.  For the density, we can take the cosmological baryon density at the redshift 
$z=0.206$ as the reference value, and describe the IGM density with a density excess parameter 
$\epsilon (\ge 1)$, $n_\mathrm{IGM}=\epsilon 4.3\times 10^{-7}$ cm$^{-3}$ \citep{spergel2003}, 
which sets the lower limit to the IGM pressure to be $\gtrsim \epsilon 5.9\times 10^{-16}$ dyne cm$^{-2}$.
In this case, if the true thermal electron pressure in B1358+305 is less than $10^{-4}\epsilon$ of 
that inferred from our upper limit, it indicates that thermal electrons do not 
play a major role in supporting the radio lobe against the external pressure. We note that this 
conclusion strongly depends on the estimate of the IGM pressure and the SZE measurement. 
Deeper X-ray observations for the pressure determination of the IGM and much sensitive SZE measurement
are required to be more conclusive.

\subsection{Comparison with X-ray Observations of Radio Galaxies in Clusters of Galaxies}  \label{sect:xray}
Finally, we discuss the X-ray observations in conjunction with the SZE in radio galaxies as 
effective means of probing the pressure of a cocoon \citep[also see][]{pfrommer2005}.
The observations by {\sl ROSAT} and {\sl Chandra} satellites have revealed that radio lobes of  radio 
galaxies centered at the cluster of galaxies generate cavities in the surface brightness distribution 
of the X-ray emitting intracluster medium 
\citep[e.g.,][]{bohringer1993,carilli1994,mcnamara2000,fabian2000,mcnamara2007}.
These are considered to be due to the pressure in the lobe overwhelming that of the surrounding hot ICM.
\citet{hardcastle2000} and \citeauthor{leahy2002} (2002), however, discovered that the ICM pressure 
inferred by the X-ray emission obtained by {\sl ROSAT} observation of bright FR-II radio galaxies 
is much higher than the minimum pressure estimated by the synchrotron luminosity in the radio band for 
large ($\gtrsim 100$\,kpc) radio galaxies.  A detailed study by \citet{leahy2001a} confirmed this 
pressure discrepancy for 3C388.  These results suggest the dominance of the other forms of pressure above 
that due to the synchrotron electrons (and magnetic field) in the radio lobes.

\citet{hardcastle2000} argued that plausible candidates for carrying the ``invisible" pressure are 
non-thermal protons, magnetic field with strength differing from that derived from the minimum 
energy condition ($\simeq B_{\rm eq}$), or low energy electrons which do not emit strong radiation 
in the observed energy bands. 
If we assume that $\xi_e$ in the internal shock is as small as that in the external shock, 
the radio lobes could have a significant amount of thermal electrons (Section\,\ref{sect:model}). 
In this case, the pressure of the thermal electrons could be one of the candidates for 
the origin of the ``invisible" pressure.
Since SZE is sensitive only to electron energy, it is a good probe of either the thermal electrons
or non-thermal electrons of low energy which do not emit high frequency emission in the observation bands. 
The observational different signature  between thermal and non-thermal SZE is the 
frequency of zero amplitude \citep[e.g.,][]{birkinshaw1999}. Specifically, the thermal SZE has the null 
frequency ($=218$\,GHz) which is almost independent of the gas temperature, 
whereas the null frequency of non-thermal SZE changes with the electron energy distribution parameters.
Hence, future multi-frequency observations that covers the null frequency of thermal SZE will be a tool to 
probe the electron contribution to the ``invisible" pressure components in the cocoon.
On the other hand, if $\xi_e$ of the internal shock is larger than that of the external shock and close to unity,  
resulting in a significant difference in the fractions of thermal and non-thermal 
electrons in the shocked shell and the radio lobes inside, SZE studies become much more difficult. 
In this case, the multi-frequency observation should be carried out over more than one region. 
For example, multi-frequency observations of the head-top region of the bow shock will be able to  
determination the contribution of mainly thermal electrons formed at the external shock.  This can be 
used to correct the observation of the central region apart from the shock-head, where 
the thermal and non-thermal SZE coexist.\footnote{For the radio galaxy 
in a galaxy cluster, another difficult problem is to distinguish the contribution from the unshocked ICM. 
In order to examine this contribution, multi-dimensional modeling of a radio galaxy embedded in the ICM 
is necessary.}
In summary, the SZE can be a probe of electron contribution to the ``invisible" pressure in a radio 
galaxy, but discrimination between thermal and non-thermal contributions requires sensitive 
multi-frequency and multi-region observations in the future. 

For the other sources of pressure, it is unlikely that the magnetic field is significantly different 
from $B_\mathrm{eq}$.  For example, X-ray observations have shown that majority of the radio galaxies 
tend to have larger particle energy than that of the magnetic field \citep[for a recent results, see][and 
references therein]{isobe2009}.  Protons, as another possible candidate for the ``invisible" pressure, on 
the other hand,  would lead to an SZE much smaller than that given by equation (\ref{eq:expect}).  The 
existence of high-energy protons should be probed in the hard X-ray and/or $\gamma$-ray regime 
\citep{scheck2002}.

\section{Conclusion}

We have reported the results of a trial observation of SZE in a giant radio galaxy B1358+305 with 
the Nobeyama 45-m telescope at 21 GHz.  By performing the imaging observation, we have obtained the most 
stringent upper limit achieved for the Compton $y$ parameter in a radio galaxy.  The obtained upper 
limit is close to the expected value derived for a low acceleration efficiency of synchrotron electrons 
at the shock (the low value $\langle \xi_e\rangle$, see Eq.\,[\ref{eq:expect}]), but detailed analysis shows that the 
obtained intensity fluctuation is likely to be caused by the excess atmospheric noise.
If we assume that the obtained intensity fluctuations in the observed region were due to the thermal 
electrons in the cocoon of B1358+305, our results are qualitatively consistent with the supersonic expansion of 
B1358+305 but for the original assumption $\langle\xi_e\rangle$ being too small.  
Alternatively, if the true pressure of thermal electrons is much lower than that derived from 
the obtained upper limit, our observation is 
not sensitive enough to derive any definitive conclusions on the 
pressure balance of B1358+305.
Since the SZE is sensitive only to electron energy, it would serve as a probe of the pressure components 
in radio galaxies as well as galaxy clusters.  Future high sensitive multi-frequency SZE observations of 
multi-regions will provide an important information about the contributions of thermal or low-energy 
non-thermal electrons to the total pressure of a radio galaxy.  This may provide a clue for disentangling 
the pressure discrepancy between the surrounding ICM derived by X-ray observations and the minimum energy 
of radio galaxies at the centers of clusters of galaxies \citep{hardcastle2000,leahy2002}.
Though we have failed to obtain a definitive SZE signal with the Nobeyama 45-m telescope, higher 
frequency observations using either a large field-of-view (e.g., $\gtrsim 5^{\prime}$), multi-beam 
receivers on large (e.g., $\gtrsim$ 30 m) single dish telescopes, or interferometers with a large number 
of small dishes (e.g., $\lesssim 50$ cm) similar to the AMiBA 
project\footnote{See {\tt http://amiba.asiaa.sinica.edu.tw/} for the AMiBA project.} might lead to 
the detection of SZE of a cocoon. The SZE would provide a new observational tool to probe the energetics of 
radio galaxies, along with the projects to detect low frequency radio emission from non-thermal low 
energy electrons like the Long Wavelength Array (LWA: \citealp{harris2005}).

\acknowledgements
We greatly appreciate N. Kuno and H. Ezawa for their help in observations with the 45-m telescope at Nobeyama Radio Observatory, and the other members of NRO that supported us during the observation.
MY especially gives her thanks to E. Komatsu for providing his programs for data analysis and 
to R. Taam for his critical reading of the manuscript.
N.S. is supported by the Sumitomo Foundation and Grant-in-Aid for scientific Research Fund (No.11640235).

{\it Facilities:} \facility{NRO(NAOJ)}.




\begin{thebibliography}{}
%
%
\bibitem[Baring et al.(1999)]{baring1999}
Baring, M. G., Ellison, D. C., Reynolds, S. P., Greiner, I. A. \& Goret, P. 1999, \apj, 513.
%
\bibitem[Begelman \& Cioffi(1989)]{begelman1989}
Begelman, M. C., \& Cioffi, D. F. 1989, \apj, 345, L21.
%
\bibitem[Birkinshaw(1999)]{birkinshaw1999}
Birkinshaw, M.  1999, Phys. Rept.,  310, 97.
%
\bibitem[Blandford \& Rees(1974)]{blandford1974}
Blandford, R. D., \& Rees, M. J. 1974, \mnras, 169, 395.
%
\bibitem[Blundell \& Rawlings(2000)]{blundell2000}
Blundell, K. M., \& Rawlings, S.  2000, \aj, 119, 1111.
%
\bibitem[B\"ohringer et al.(1993)]{bohringer1993}
B\"ohringer, H., Voges, W., Fabian, A. C., Edge, A., C., Neumann, D. M.  1993, \mnras, 264, L25.
%
\bibitem[Burns, Norman, \& Clarke(1991)]{burns1991}
Burns, J. O., Norman, M. L., \& Clarke, D. A.  1991, Science, 253, 522.
%
\bibitem[Carlli, Perley, \& Harris(1994)]{carilli1994}
Carilli, C, L., Perley, R. A., \& Harris, D. E. 1994, \mnras, 270, 130.
%
\bibitem[Carlstrom et al.(2001)]{carlstrom2001}
Carlstrom, J. E., Joy, M., Grego, L., et al.  2001"Constructing the Universe with Clusters of 
Galaxies", IAP Conf.  eds. F. Durret \& G. Gerbal (arXiv:astro-ph/0103480).
%
\bibitem[Clarke, Norman \& Burns(1986)]{clarke1986}
Clarke, D. A., Norman, M. L., \& Burns, J, O.  1986, \apj, 311, L63.
%
\bibitem[Colafrancesco(2008)]{colafrancesco2008}
Colafrancesco, S. 2008, \mnras, 385, 2041.
%
\bibitem[Croston et al.(2005)]{croston2005}
Croston, J. H. et al. 2005, \apj, 626, 733.
%
\bibitem[Croston et al.(2009)]{croston2009}
Croston, J. H. et al. 2009, \mnras, 395, 1999.
%
\bibitem[En\ss lin \& Kaiser(2000)]{ensslin2000}
En\ss lin, T. A., \& Kaiser, C. R.  2000, \aap, 360, 417.
%
\bibitem[Fabian et al.(2000)]{fabian2000}
Fabian, A. C., Sanders, J. S., Ettori, S., et al.  2000, \mnras, 318, L65.
%
\bibitem[Fanaroff \& Riley(1974)]{fanaroff1974}
Fanaroff, B. L., \& Riley, J. M.  1974, \mnras, 167, 31.
%
\bibitem[Franceschini et al.(1989)]{franceschini89}
Franceschini, A., Toffolatti, L., Danese, L., \& De Zotti, G.  1989, \apj, 344, 35.
%
\bibitem[Grandi et al.(2003)]{grandi2003}
Grandi, P. et al. 2003,\apj. 586, 123.
%
\bibitem[Heinz, Reynolds, \& Begelman(1998)]{heinz1998}
Heinz, S., Reynolds, C. S., \& Begelman, M. C.  1998, \mnras, 501, 126.
%
\bibitem[Hardcastle \& Worrall(2000)]{hardcastle2000}
Hardcastle, M. J., \& Worrall, D. M.  2000, \mnras, 319, 562.
%
\bibitem[Hardcastle, Birkinshaw, and Worrall(2001)]{hardcastle2001}
Hardcastle, M. J., Birkinshaw, M., \& Worrall, D. M.  2001, \mnras, 323, L17.
%
\bibitem[Hardcastle \& Looney(2008)]{hardcastle2008}
Hardcastle, M. J. \& Looney, 2008, \mnras, 388, 176.
%
\bibitem[Harris, Carilli, \& Perley(1994)]{harris1994}
Harris, D. E., Carilli, C. L., \& Perley, R. A.  1994, Nature, 367, 713.
%
\bibitem[Harris, Leighly, \& Leahy(1998)]{harris1998}
Harris, D. E., Leighly, K. M., \& Leahy, J. P. 1998, \apjl, 499, L149.
%
\bibitem[Harris et al.(1999)]{harris1999}
Harris, D. E., Hjorth, J., Sadun, A. C., Silverman, J. D., Vestergaard, M. 1999, \mnras, 518, 213.
%
\bibitem[Harris et al.(2000)]{harris2000}
Harris, D. E., Nulsen, P. E. J., Ponman, T. J.,  et al.  2000, \apj, 530, L81.
%
\bibitem[Harris(2005)]{harris2005}
Harris, D. E. F, 2005, proceedings of "Clark Lake to the Long Wavelength Array: Bill 
Erickson's Radio Science", ASP Conf. Ser, vol. 345, 254.
%
\bibitem[Isobe et al.(2006)]{isobe2006}
Isobe, N., et al. 2006, \apj, 645, 256.
%
\bibitem[Isobe et al.(2009)]{isobe2009}
Isobe, N. et al. 2009, \apj, 706, 454.
%
\bibitem[Ito et al.(2008)]{ito2008}
Ito, H., Kino, M., Kawakau, N., Isobe N. \& Yamada, S. 2008, \apj, 685, 828.
%
\bibitem[Kaiser \& Alexander(1999)]{kaiser1999}
Kaiser, C., \& Alexander, P. 1999, \mnras, 305, 707.
%
%
\bibitem[Komatsu et al.(2001)]{komatsu2001}
Komatsu, E., Matsuo, H., Kitayama, T., et al.  2001, PASJ, 53, 57.
%
\bibitem[Konar et al.(2009)]{konar2009}
Konar, C., Hardcastle, M. J., Croston, J. H. \& Saikia, D. J. 2009, \mnras, 400, 480.
%
\bibitem[Kuno(1993)]{kuno1993}
Kuno, N.  1993, PhD Thesis, Tohoku University.
%
\bibitem[Leahy \& Gizani(2001)]{leahy2001a}
Leahy, J. P., \& Gizani, N. A. B. 2001, \apj, 555, 709.
%
\bibitem[Leahy \& Gizani(2002)]{leahy2002}
Leahy, J. P., \& Gizani, N. A. B. 2002, NewA Rev., 46, 117.
%
\bibitem[Lind et al.(1989)]{lind1989}
Lind, K. R., Payne, D. G., Meier, D. L., \& Blandford, R. D. 1989, \apj, 344, 89.
%
%
\bibitem[Loeb \& Waxman(2000)]{loeb2000}
Loeb, A.\& Waxman, E. 2000, \nat, 405, 156.
%
\bibitem[Loken et al.(1992)]{loken1992}
Loken, C., Burns, J. O., Clarke, \& Norman, M. L.  1992, ApJ, 392, 54.
%
\bibitem[McKinnon et al.(1991)]{mckinnon91}
McKinnon M.M., Owen F.N., \& Eilek J.A. 1991, \aj, 101, 2026.
%
\bibitem[McNamara et al.(2000)]{mcnamara2000}
McNamara, B. R., Wise, M., Nulsen, P. E., et al.  2000, \apjl, 534, L135.
%
\bibitem[McNamara \& Nulsen(2007)]{mcnamara2007}
McNamara, B. R. \& Nulsen, P. E. J. 2007, \araa, 45, 117.
%
\bibitem[Moffet(1975)]{moffet1975}
Moffet, A. T. 1975, "Galaxies and the Universe", ed. Sandage, A., Sandage, M. \& Kristian, J., 
the Chicago University Press.
%
%
\bibitem[Nath(1995)]{nath1995}
Nath, B. B. 1995, \mnras, 274, 208.
%
\bibitem[Ott et al.(1994)]{ott1994}
Ott, M., Witzel, A., Quirrenbach, A., et al. 1994, \aap, 284, 331.
%
\bibitem[Parma et al.(1996)]{parma1996}
Parma, P., de Ruiter, H. R., Mack, K.-H.,  et al. 1996, \aap, 311, 49.
%
\bibitem[Pfrommer, En\ss\,lin \& Sarazin(2005)]{pfrommer2005}
Pfrommer, C., En\ss\,lin, T. A. \& Sarazin, C. L. 2005, \aap, 430, 799.
%
\bibitem[Rephaeli(2002)]{rephaeli2002}
Rephaeli, Y.  2002, Space Science Reviews, 100,  61.
%
%
\bibitem[Saripalli et al.(1996)]{saripalli1996}
Saripalli, L., Mack, K.-H., Klein, U., Strom, R., Singal, A. K.  1996, \aap, 306, 708.
%
\bibitem[Sarazin(1999)]{sarazin1999}
Sarazin, C. L. 1999, \apj, 520, 529.
%
\bibitem[Scheck et al.(2002)]{scheck2002}
Scheck, L., Aloy, M. A., Mart\'{i}, J. M., G\'{o}mez, J. L. \& M\"{u}ller, E., 2002, \mnras, 331, 615.
%
\bibitem[Scheuer(1974)]{scheuer1974}
Scheuer, P. A. G. 1974, \mnras, 166, 513.
%
\bibitem[Sieber, Haslam, \& Salter(1979)]{sieber1979}
Sieber, W., Haslam, C. G., \& Salter, C. J. 1979, \aap, 74, 361.
%
\bibitem[Spergel et al.(2003)]{spergel2003}
Spergel, D. N. et al. \apjs, 148, 175.
%
\bibitem[Sturner et al.(1997)]{sturner1997}
Sturner, S. J., Skibo, J. G., Dermer, C. D., \& Mattox, J. R. 1997, \apj, 490, 619.
%
\bibitem[Tashiro et al.(1998)]{tashiro1998}
Tashiro, M. et al. 1998, \apj, 499, 713.
%
\bibitem[Wardle et al.(1998)]{wardle1998}
Wardle, J. F. C., Homan, D. C., Ojha, R. \& Roberts, D. H. 1998, \nat, 395, 457.
%
\bibitem[Wilson, Young, \& Shopbell(2000)]{wilson2000}
Wilson, A. S., Young, A. J., \& Shopbell, P. L.  2000, \apj, 544, L27.
%
\bibitem[Wilson, Young, \& Shopbell(2001)]{wilson2001}
Wilson, A. S., Young, A. J., \& Shopbell, P. L.  2001, \apj, 547, 740.
%
\bibitem[Yamada et al.(1999)]{yamada1999}
Yamada, M., Sugiyama, N., \& Silk, J. \ 1999, \apj, 522, 66.
%
\bibitem[Yamada \& Fujita(2001)]{yamada2001}
Yamada, M., \& Fujita, Y.  2001, \apj, 553, 145.
%
\bibitem[Zel'dovich \& Sunyaev(1969)]{zeldovich1969}
Zel'dovich, Ya. B.,  \& Sunyaev, R. A.  1969, \apjs, 4, 301.
%
\end{thebibliography}
\end{document}